\documentclass{aastex631}

\usepackage{bm}
\usepackage{cleveref}

\usepackage{fontawesome5}
\newcommand{\github}[1]{\href{#1}{\faGithubSquare}}
\newcommand{\githublink}{\github{https://github.com/harrydesmond/GalaxySpinAnisotropy}}

\begin{document}

\title{Symmetry in Hyper Suprime-Cam galaxy spin directions}

\author[0000-0002-0986-314X]{Richard Stiskalek}
\affiliation{Astrophysics, University of Oxford, Denys Wilkinson Building, Keble Road, Oxford, OX1 3RH, UK}

\author[0000-0003-0685-9791]{Harry Desmond}
\affiliation{Institute of Cosmology \& Gravitation, University of Portsmouth, Dennis Sciama Building, Portsmouth, PO1 3FX, UK}

\correspondingauthor{Richard Stiskalek, Harry Desmond}
\email{richard.stiskalek@physics.ox.ac.uk, harry.desmond@port.ac.uk}

\begin{abstract}

We perform a Bayesian analysis of anisotropy in binary galaxy spin directions in the Hyper-Suprime Cam Data Release 3 catalogue, in response to a recent claim that it exhibits a dipole~\citep{Shamir_2024}. We find no significant evidence for anisotropy, or for a direction-independent spin probability that differs from 0.5. These results are unchanged allowing for a quadrupole or simply searching for a fixed anisotropy between any two hemispheres, and the Bayes factor indicates decisive evidence for the isotropic model. Our principled method contrasts with the statistic employed by~\citet{Shamir_2024}, which lacks a strong theoretical foundation. Our code is available at \githublink.

\end{abstract}


\section{Introduction}\label{sec:introduction}
    
The Universe is thought to be homogeneous and isotropic on large scales, in accordance with the cosmological principle. However, several observations such as the cosmic microwave background anomalies, bulk flows, and discrepancies between the cosmic microwave background and distant matter rest frames, point to possible anisotropies (e.g.,~\citealt{CF4,EB_1}). There have been claims for anisotropy in the directions of galaxy spins, forming a dipole axis that would violate large-scale anisotropy (e.g.,~\citealt{Longo, McAdam_Shamir_2023_GAN}). Most recently,~\cite{Shamir_2024} claims a more than $3\sigma$ detection of such a dipole in spins derived from Hyper Suprime-Cam Data Release 3 (HSC DR3;~\citealt{HSC_DR3}).~\citet{Shamir_2024} also claims a monopole in spin probability that is inconsistent with 0.5, with galaxies rotating opposite to the Milky Way (as seen from Earth) significantly more common than those rotating in the same direction.

The claims of anisotropy prior to \citet{Shamir_2024} were recently revisited by~\cite{Patel_2024} who found no significant evidence for a dipole or a monopole differing from 0.5 in any dataset publicly available at that time. This was shown through both a standard Bayesian and frequentist analysis. The discrepancy with previous analyses was found to be the poorly motivated statistics that they employed. Here we adapt the framework of~\citeauthor{Patel_2024} to the new HSC data to show that this also does not indicate an anomalous monopole, dipole or quadrupole.

\section{Methodology}\label{sec:methodology}

We take the HSC DR3 data which matches that used by~\citet{Shamir_2024}. We assume that these spin assignments are correct; a direction-dependent bias in the assignment would be much more likely to introduce a spurious dipole than spuriously remove a true one. The catalogue is illustrated in the left panel of~\Cref{fig:galaxy_spin_distribution}. We follow the methodology of~\citeauthor{Patel_2024}. We denote the spin direction of the $i$\textsuperscript{th} galaxy relative to the Milky Way as $s_i$, which can be either Z-wise or S-wise. The likelihood of Z-wise spin is
\begin{equation}
    \mathcal{L}(s_i | M, D, \hat{\bm{d}}, Q, \hat{\bm{q}}_1, \hat{\bm{q}}_2)
    = M
    + D\left(\hat{\bm{d}} \cdot \hat{\bm{n}}_i\right)
    + Q\left(\hat{\bm{q}}_1 \cdot \hat{\bm{n}}_i ~ \hat{\bm{q}}_2 \cdot \hat{\bm{n}}_i - \frac{1}{3}\hat{\bm{q}}_1\cdot\hat{\bm{q}}_2\right)
\end{equation}
where $M$ is the monopole, $D$ the dipole magnitude and $\hat{\bm{d}}$ is a unit vector pointing in the direction of the dipole. $Q$ is the strength of a possible quadrupole with unit axes $\hat{\bm{q}}_1$ and $\hat{\bm{q}}_2$. $\hat{\bm{n}}_i$ is the unit vector in the direction of the galaxy. We use uniform priors on $M$, $D$ and $Q$ and on the area element of all unit vectors.  The likelihood of S-wise spin is $1 - \mathcal{L}$, and we assume that all galaxies are independent such that the dataset likelihood is $\prod_i \mathcal{L}_i$. Isotropy corresponds to $D = 0$ and $Q = 0$, and an overall balance between Z-wise and S-wise spins corresponds to $M = 0.5$.

We upgrade the code of~\citeauthor{Patel_2024} to \texttt{JAX}\footnote{\url{https://jax.readthedocs.io/en/latest/}}, sampling the posterior using the No U-Turns Sampler~\citep{Hoffman_2011} method of Hamiltonian Monte Carlo algorithm (HMC) implemented in \texttt{NumPyro}~\citep{Phan_2019}. We remove burn-in and use sufficient steps for the Gelman--Rubin statistic to be 1 to within $10^{-3}$~\citep{Gelman_1992}.

\section{Results and Discussion}\label{sec:conclusion}

\begin{figure}[ht]
    \centering
        \raisebox{0.5\height}{\includegraphics[width=0.49\textwidth]{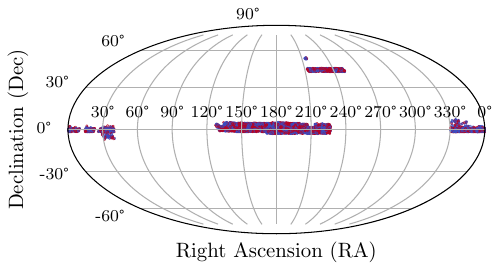}}
    \hfill
        \includegraphics[width=0.49\textwidth]{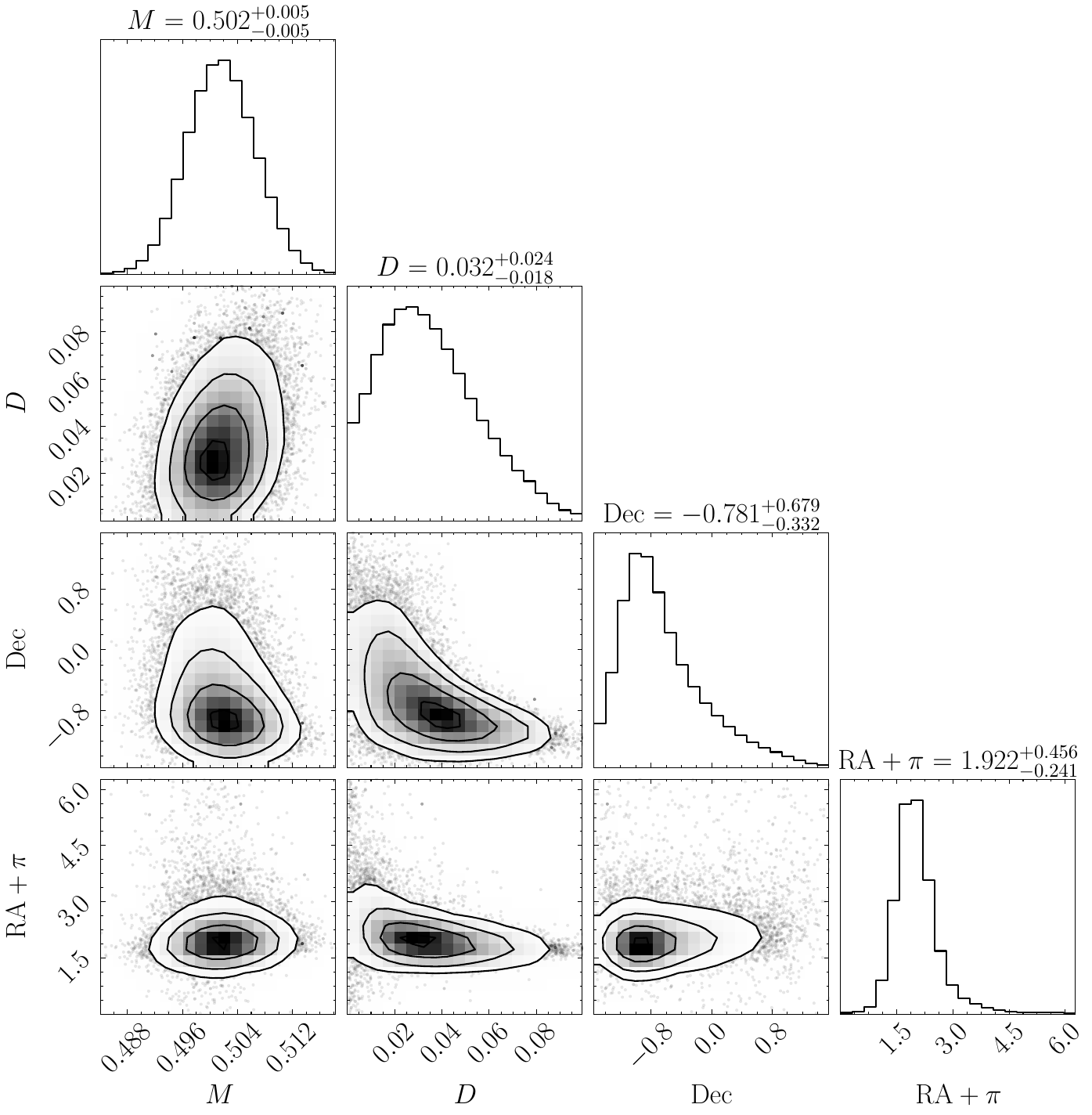}
    \caption{\emph{Left:} Sky distribution of the HSC DR3 galaxies in equatorial Mollweide projection, with galaxies spinning S-wise in blue and Z-wise in red. \emph{Right:} Corner plot from the monopole-plus-dipole inference, indicating a dipole magnitude $D$ consistent with 0 and a monopole $M$ consistent with 0.5.}
    \label{fig:galaxy_spin_distribution}
\end{figure}

We first allow a monopole and dipole, showing the posterior in the right panel of ~\Cref{fig:galaxy_spin_distribution}. We find that $M$ is consistent with $0.5$, indicating no preference for one spin direction over the other, and that $D$ is consistent with $0$ to within $2\sigma$, indicating no significant dipole. That $M\approx0.5$ is unsurprising given that the average spin in the sample is 0.499, and that $D\approx0$ is unsurprising given the poor sky coverage of the HSC data. This makes the constraints on $D$ a few times weaker than those of the other datasets studied in~\citeauthor{Patel_2024}. We then run the monopole--dipole--quadrupole inference, finding the same results and that $Q$ is consistent with $0$. Following~\citet{Shamir_2024} we also try splitting the data into the redshift ranges $0<z<0.1$ and $0.1<z<0.2$, finding near-identical results in both cases. We also investigate the ``hemisphere anisotropy'' model of~\citeauthor{Patel_2024} which neglects the $\cos\theta$ dependence of the dipole, finding the anisotropy parameter $A$ to be consistent with 0 within $2\sigma$.

Finally, we compute the Bayes factor describing the relative probability of the monopole--only and monopole--dipole models, i.e. the preference for adding a dipole. As the Bayesian evidence is prior-dependent we try two ranges for the uniform prior on $D$: $0$ to $0.1$ and $0$ to $0.5$ (both fully enclosing the posterior in all cases). We use the \texttt{harmonic} package~\citep{Polanska_2024} to compute it directly from the HMC chain. We find $\log_{10}$ evidence ratios in favour of the monopole-only model of 5.71 and 4.61 for the looser and tighter $D$ priors, respectively. Even in the latter case this corresponds to a Bayes factor of $40,738$ in favour of the monopole-only model, which is ``decisive'' on the Jeffreys' scale.

We conclude that there is no evidence for an anisotropy of any kind in the spin directions of the HSC DR3 dataset, just as there is not in any other. That the opposite is found in~\citet{Shamir_2024} is attributable to the use of an off-$\chi^2$ statistic which relies on assumptions that may not be fully justified, as detailed in Sec. 4.3 of~\citeauthor{Patel_2024}.

\section*{Data availability}

Our analysis code is available at \githublink~and \texttt{Zenodo}~\citep{harry_desmond_2024_14036930}. The HSC DR3 data is available at \url{https://people.cs.ksu.edu/~lshamir/data/asymmetry_hsc/}.

\section*{Acknowledgements}

We thank Dhruva Patel for contributions to the paper on which our analysis is based. RS acknowledges financial support from STFC Grant No. ST/X508664/1 and the Snell Exhibition of Balliol College, Oxford. HD is supported by a Royal Society University Research Fellowship (grant no. 211046). For the purpose of open access, we have applied a Creative Commons Attribution (CC BY) licence to any Author Accepted Manuscript version arising.

\bibliography{ref.bib}{}
\bibliographystyle{aasjournal}

\end{document}